**Simulation study of various factors affecting the performance of Vertical Organic Field-Effect Transistors**

*Ramesh Singh Bisht, and Pramod Kumar**

Ramesh Singh Bisht, Pramod Kumar
Department of Physics, Indian Institute of Technology Bombay, Maharashtra, 400076, India
E-mail: pramod_k@iitb.ac.in

Vertical field effect transistors (VOFETs) can offer short channel architecture which can further enhance the performance at low operating voltages which makes it more viable for organic electronics applications. VOFETs can be prepared with low-cost techniques which reduce the high processing costs and can also operate at high current density and relatively higher frequencies. To further improve the performance, high current density, and operating frequency the physics of charge carrier transport should be understood well with the simulation. The main problem with VOFET is the high off-current which is inevitable due to conduction from source to drain contact. There have been many efforts in reducing the off-state current by the addition of an insulating layer on top of the source electrode, which further increases the complexity and cost of processing. Simulations based on device geometry, contact barriers, and organic semiconductor parameters are carried out to study the charge carrier transport in VOFET. The simulation results show that the most important factor to enhance the performance is the device geometry or architecture, which requires a specific fill factor, a ratio between the exposed gate dielectric, and the total length with the source electrode. Optimized VOFET architecture is then simulated for variation in contact barrier and semiconductor parameters, which show some enhancement in performance but also a rise in off-state current density.

## 1. Introduction

Achieving good performance and higher switching speeds in organic field effect transistors (OFETs) are required for the development of devices comparable to silicon technology. Organic field effect transistors provide a low-cost alternative to field effect transistors for the development of flexible electronics.[1–6] Generally organic semiconductors (OSCs) have low mobility and high traps due to inherent disorder which can hinder the performance and high frequency response of the devices.[7–10] There are three ways in which researchers are working to rectify this problem, one is to develop new OSC molecules and polymers, second by finding suitable morphology of the organic layer for better charge carrier conduction, and third by the development of new device architectures.[11–17] An addition to the device architecture is the vertical organic field-effect transistors (VOFETs) in which charge carrier transport occurs along the vertical source and drain contact and the perforations on the source electrode allow the gate field to be applied along the vertical direction which enhances charge carrier transport by accumulation in those areas. Initial VOFETs type devices were introduced by various research groups across the globe[18–20] and later discontinuous source electrode was introduced by Y. Yang et al.[21], here the aim was to increase the device current despite low charge carrier mobility and traps due to inherent disorder. This architecture was later modified using large-scale gate perforations on the source electrode which is well known as Patterned Source-VOFET (PS-VOFET) designed and fabricated by B. Sasson et al.[16,22,23] Unlike the traditional lateral geometry of OFET, where the gate field is applied perpendicular to the source-drain field, in VOFETs the flow of charge carrier occurs in the vertical direction and the gate field is also applied in the same direction. The channel length in lateral OFETs is defined as the horizontal separation between source to drain electrodes and it is limited to a micrometer scale due to the limitation of the lithography techniques.[24] VOFET offers the possibility of reduction of the channel length in the submicron range and up to a few tens of nanometer ranges since the channel is along the vertical direction and depends only on the thickness of the semiconductor layer deposited on top of the source/gate-dielectric. Reducing the channel length of the device enhances the current density at low operating voltages and also provides operations at switching speed which would benefit current and future flexible devices like radio-frequency identification tags (RFIDs),[25–28] flexible displays,[29] and bio-sensors.[30,31] Interest in this novel geometry for electronics applications and the study of the underlying physics of charge carrier transport has been going on for the past many years.[32,33] The charge carrier transport in VOFETs occurs by the formation of a channel by applying the gate field which accumulates the charge carriers in the perforations which increases the carrier concentration in these regions and the source drain field can easily extract the charge carriers.[34] The traditional OFET performance is always inhibited by the longer charge carrier path along the channel which results in low operational frequency. This drawback can be sorted out by using VOFET geometry with comparatively less charge carrier path to cover from one electrode (source) to another (drain) with the extra advantage of low operational voltages (< 5 V). Relatively faster switching VOFET devices have been experimentally realized with a high On/Off ratio (~ $10^4$) and high current density and are also analyzed by simulation techniques by M. Greenman et al.[35, 36] In this work there have been many attempts to reduce the off-state current by adding extra non-insulating layers on the source electrode and also by sandwiching an extra OSC layer between the dielectric and source electrode, which increases the processing cost and complexity of fabrication. To find a suitable solution to these problems, first, the charge carrier transport in VOFETs must be understood so that some ambiguities which are remaining regarding its working can be resolved.

Here we report the simulation study of the charge carrier transport in VOFET using COMSOL Multiphysics software to find out various factors which can affect charge carrier movement in the OSC channel and hence reduce the off-state current and enhance the performance. Three main factors were investigated for this study viz. device geometry or architecture, contact barriers, and OSC properties. Simulations of all these factors can throw light on the physics of charge carrier transport in VOFETs and also suggests modification under which a good device performance can be achieved. The simulation uses the drift-diffusion physics equations for charge carriers (electrons and holes) and a 2D device design which exhibits perforated source electrodes. The design of the simulated VOFETs consists of a perforated source electrode on top of the dielectric gate layer and a semiconductor layer of 100 nm which is sandwiched between the top drain and the perforated source electrode. Various factors which can affect current density ($J$), On/Off ratio, and subthreshold swing (SS) were systematically identified, starting from the basic geometrical modification in the structure of the device. The changes in basic device architecture include the gate width, source width, and source height changes. The second factor includes different source electrode materials with different injection barriers, and the third factor is the charge carrier mobility and free charge carrier concentration in the OSC layer. These parameters were changed in the simulation to understand the physics of charge carrier transport in VOFETs, so that a recipe for high-performance devices with low off-state current can be identified.

## 2. Simulation and results

The simulation work in this paper basically is related to the characteristic changes in the performance of patterned source-vertical field-effect transistors (PS-VOFETs). The VOFET device is comprised of three electrodes, a dielectric layer and a semiconducting layer, namely the gate electrode on the dielectric, located at the bottom of the structure, the source electrode, perforated or discontinuous patterned type, located on top of the gate dielectric ($SiO_2$), OSC layer on top of the source electrode and dielectric layer (perforated region), and the drain electrode on top of the device. The OSC material is sandwiched between the top drain electrode and gate dielectric region alongside the source electrode at the bottom. OSCs provide an active channel for the conduction of charge carriers when a gate field is applied for the accumulation of charge carriers in the perforated regions. The structure of the simulated VOFET device is illustrated in **Figure 1** which demonstrates the vertically stacked layers of various components of the device.

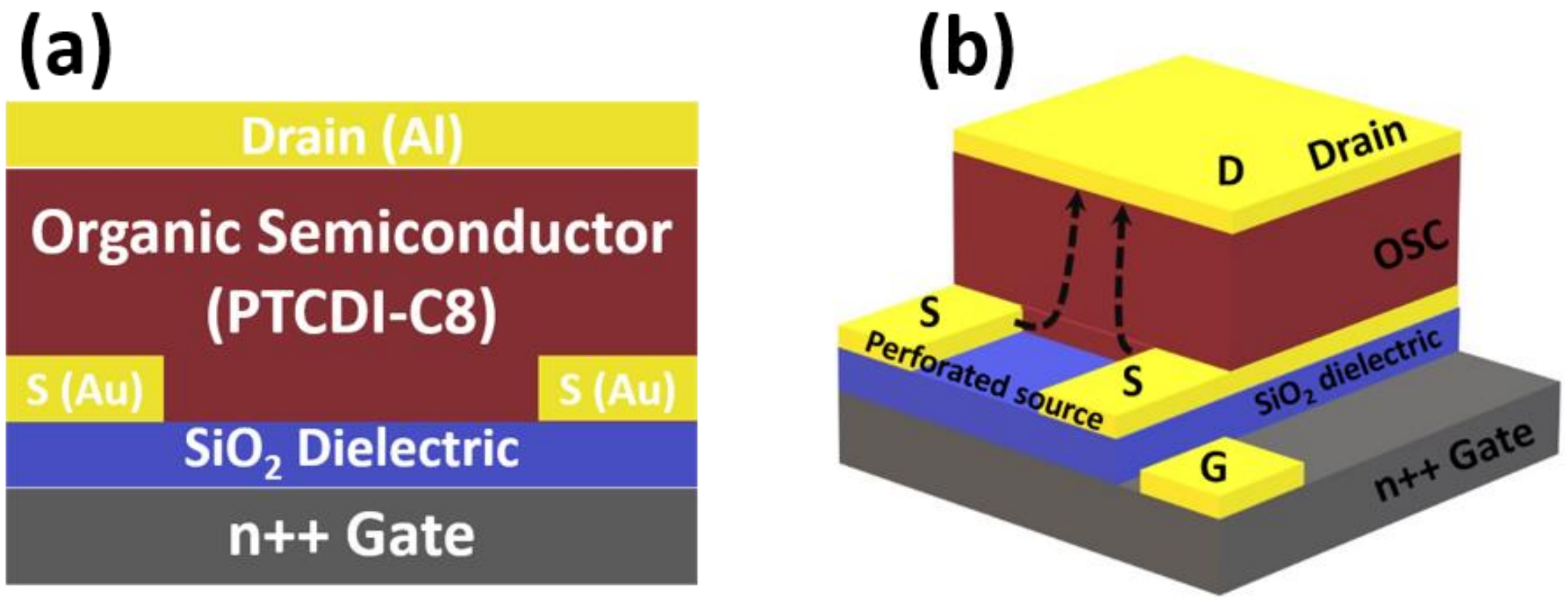

**Figure 1.** Illustration of single PS-VOFET in (a) 2D, and (b) 3D; the source electrode (S) is discontinuous and charge carrier conduction takes place between the source and drain contacts when a gate field is applied for charge carrier accumulation as shown by arrows in the 3D model.

Simulations are done with COMSOL Multiphysics software using two-dimensional modeling, and the 'semiconductor (semi) module physics' package for the design of VOFET and for finding out various factors affecting its performance. **Figure 2(a)** shows the image of a designed VOFET with all the layers and **Figure 2(b)** is the model with dimensions in the COMSOL Multiphysics simulation software. The structure contains a gate electrode which is highly doped silicon (*n*-type) and gate dielectric material- $SiO_2$ of a thickness of 30 nm on top of it, which is represented as the horizontal red bottom line at 0 nm height. The perforated source electrode (Au) is having 10 nm height ($h_s$) on top of the gate dielectric with a perforation region of 20 nm which is defined as gate width ($w_g$). The source electrode is represented as vertical and horizontal green lines as shown in **Figure 2(b)**. The OSC chosen here is *n*-type PCTDI-C8 which has a thickness of 100 nm. Finally, the drain electrode (Al) is placed on top of the OSC layer, represented as the top blue line in **Figure 2(b)**. The in-build drift-diffusion mechanism of the semiconductor module of COMSOL is applied to this device (without any additional changes) as follows:

$$\rho + = q(p - n + N_d^+ - N_a^-) \qquad (1)$$

Equation 1 refers to the total charge concentration ($\rho +$) in the channel of the device, where $q$, $p$, and $n$ denote the elementary charge, hole density, and electron density respectively. $N_d^+$ and $N_a^-$ are the donor atom and the acceptor atom concentrations respectively.

$$\nabla \cdot J_n = 0, \qquad \nabla \cdot J_p = 0 \qquad (2)$$

Equation 2 represents the continuity equations for respective charge carriers (electrons and holes) and $J$ denotes the current density.

$$J_n = qn\mu_n \nabla E_c + qD_n \nabla n - qnD_n \nabla\, ln(N_c) + qnD_{n,th} \nabla ln(T) \qquad (3)$$

$$J_p = qp\mu_p \nabla E_v - qD_p \nabla p + qpD_p \nabla\, ln(N_v) - qpD_{p,th} \nabla ln(T) \qquad (4)$$

In Equations 3 and 4, the charge carrier current densities are shown which is simulated in this module, where $\mu_n$, $\mu_p$, $D_n$ and $D_p$ are the mobility and diffusion coefficients for electrons and holes respectively. $E_c$, $E_v$, $N_c$ and $N_v$ are the charge carrier energy levels and density of states at conduction and valence bands respectively. $D_{n,th}$, $D_{p,th}$, and $T$ represent the electron thermal diffusion coefficient, hole thermal diffusion coefficient, and temperature, respectively.

$$E_c = -(V + \chi_0) \quad , \qquad E_v = -(V + \chi_0 + E_{g,0}) \qquad (5)$$

Equation 4 shows the energy of the charge carriers at conduction and valence band edges respectively. $V$, $\chi_0$ and $E_{g,0}$ are the applied potential, electron affinity, and band gap at equilibrium in the semiconductor channel, respectively.

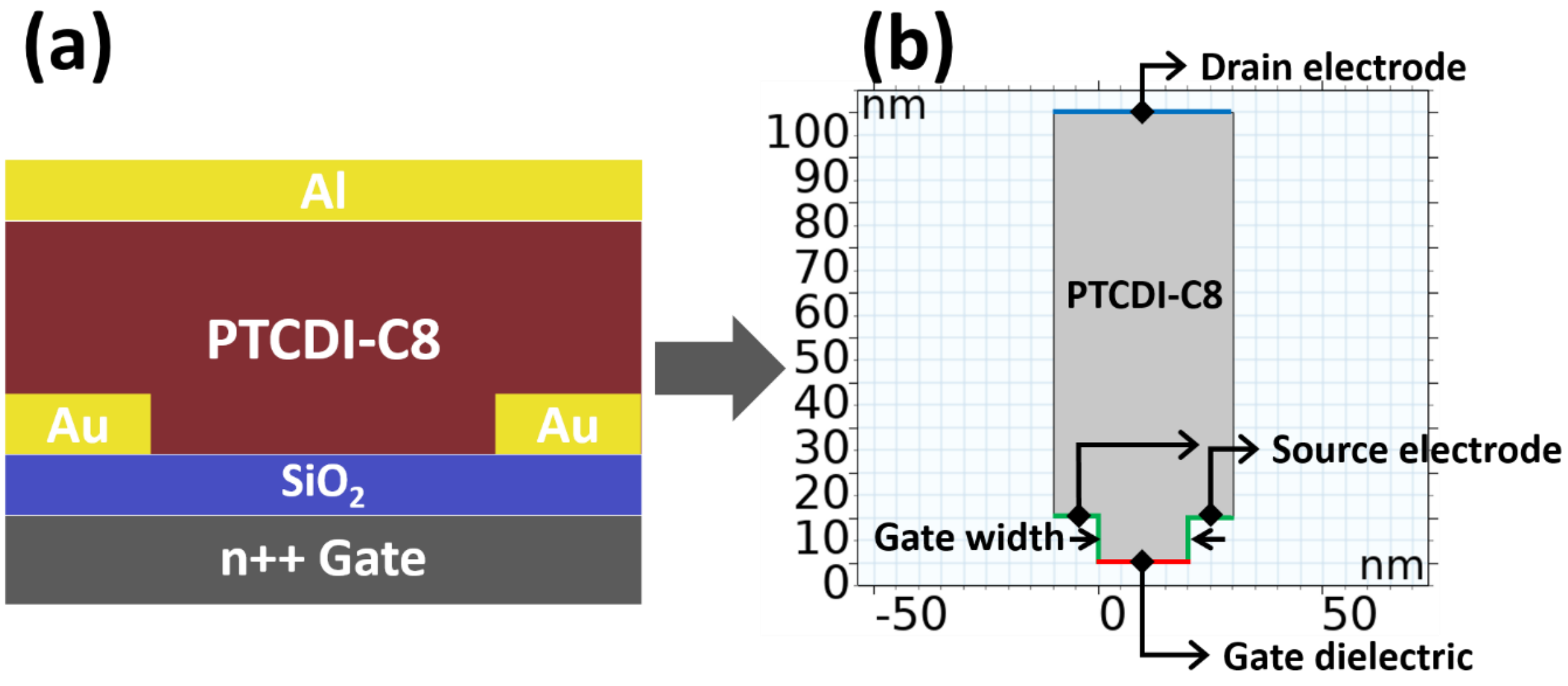

**Figure 2. (a)** VOFET architecture with various layers of its components. **(b)** Corresponding model of same VOFET in COMSOL simulation software with gate width of 20 nm.

## 2.1. Effect of VOFET geometry

The VOFET structure can be modified in three structural ways viz. the source electrode perforation size/gate opening, the thickness of the source electrode, and the width of the source electrode. These three factors are systematically simulated for understanding the physics of charge carrier transport and also to improve the device performance and off-state current.

### *2.1.1. Source electrode perforation size/gate width/perforation region*

The basic operating principle of VOFET lies in the size of gate width or source perforation, the perforation regions are responsible for the accumulation of charge carriers when a gate field is applied, and due to the accumulation, these regions provide a less resistive path for conduction from source to drain or vice versa. Five different gate widths/source perforations are chosen for simulation viz. 20 nm, 50 nm, 100 nm, 150 nm, and 180 nm to observe its effect on VOFET performance. The numerical simulations of different devices are done with all the above gate opening/gate width ($w_g$) for finding out the $J_D$-$V_G$ curves. All other device parameters are kept the same for all the devices i.e. source height, $h_s$=10 nm, source width=10 nm, OSC thickness of 100 nm, N-type donor concentration/free charge carriers, $N_{D0}$ = 1×10$^{12}$ cm$^{-3}$, source voltage, $V_S$ = 0 V, drain voltage $V_D$ = 0.1 V, charge injection barrier height, $\Phi_B$ = 0.44 eV, gate oxide thickness, $d_{ins}$ = 30 nm, oxide relative permittivity, $\varepsilon_{rins}$ = 3.9, and source metal work function, $\Phi_S$ = 5.0 eV. The simulations are done for an N-type OSC viz. N, N′-dioctyl-3,4,9,10-perylenedicarboximide (PTCDI-C8) with all the parameters mentioned in **Table 1.**[37,38] These parameter values are used in all simulations throughout this work.

**Table 1.** Basic semiconductor properties of PTCDI-C8 used in the simulations.

| Relative permittivity, $\varepsilon_r$ | Band gap, $E_{g0}$ [eV] | Electron affinity, $\chi_0$ [eV] | Effective density of states, val. band, $N_v$ [cm$^{-3}$] | Effective density of states, cond. band, $N_c$ [cm$^{-3}$] | Electron mobility, $\mu_n$ [cm$^2$ V$^{-1}$ s$^{-1}$] | Hole mobility, $\mu_p$ [cm$^2$ V$^{-1}$ s$^{-1}$] |
|---|---|---|---|---|---|---|
| 3 | 2 | 4.3 | $1\times10^{18}$ | $1\times10^{20}$ | $1\times10^{-5}$ | $1\times10^{-5}$ |

During the COMSOL simulation, various parameters can also be monitored which include charge carrier concentration profile and electric field lines in the simulated device, so with an n-type OSC, the electron concentration (log scale), log(n) is obtained and are shown in **Figure 3(a-c)**. The figures show different opening widths of source electrode/gate width (20 nm, 50 nm, and 150 nm) for comparison where the gate field and drain electric field vectors are also shown with the help of arrows at $V_G$=5 V and $V_D$=0.1 V. The effect of gate width is clearly visible by comparing Figure 3(a), (b) and (c), where it can be seen that as the gate width increases, the gate field interact in a larger region and can penetrate deeper in the channel in case higher (150 nm) perforation size which suggests better control of device current with gate field. The accumulated charge carrier density also shows deeper penetration into the device by which it can participate in the conduction. **Figure 3(d)** shows the charge carrier distribution throughout the OSC from gate to drain end at $w_g$=100 nm, $V_G$=5 V, and $V_D$=2 V. The channel formation is visible in the figure from gate dielectric/OSC interface to the drain electrode, which demonstrates the working of a typical PS-VOFET. Comparing it with Figure 3(a) which shows that the accumulation channel does not penetrate the channel length and is confined to a small region, hence it cannot participate much in the conduction mechanism.

The transfer current density outcomes of all the simulated PTCDI-C8-based VOFETs with different gate widths/source perforations are shown in **Figure 4**. The figure shows that the drain current density $J_D$ are decreasing on increasing the gate width at a particular $V_{DS}$ and at the same time On/Off ratios are increasing due to the effect of fill factor (FF) (which is defined as a gate to total width ratio of the device which also includes source electrode). The low On/Off ratio in the case of smaller gate width can also be corroborated by the fact that the gate electric field penetration is limited in such devices as seen in Figure 3(a). The high off-state current density at smaller source perforations (at a voltage where there is no accumulation $V_G$= -2 V), is mostly from the upper part of the source electrode and the drain electrode. The higher current density in both off and on-states of smaller source perforations can be correlated with the FF of the devices. For 20 nm source perforation, the current is highest in both on and off-states, but the gate effect is also lowest as seen in Figure 4. The performance of VOFET can be compared with the values of two basic parameters which are On/Off ratios and SS [SS = $\Delta V_G/\Delta(\log J_D)$ ]. The best-performing VOFETs should show large On/Off ratios and low SS.

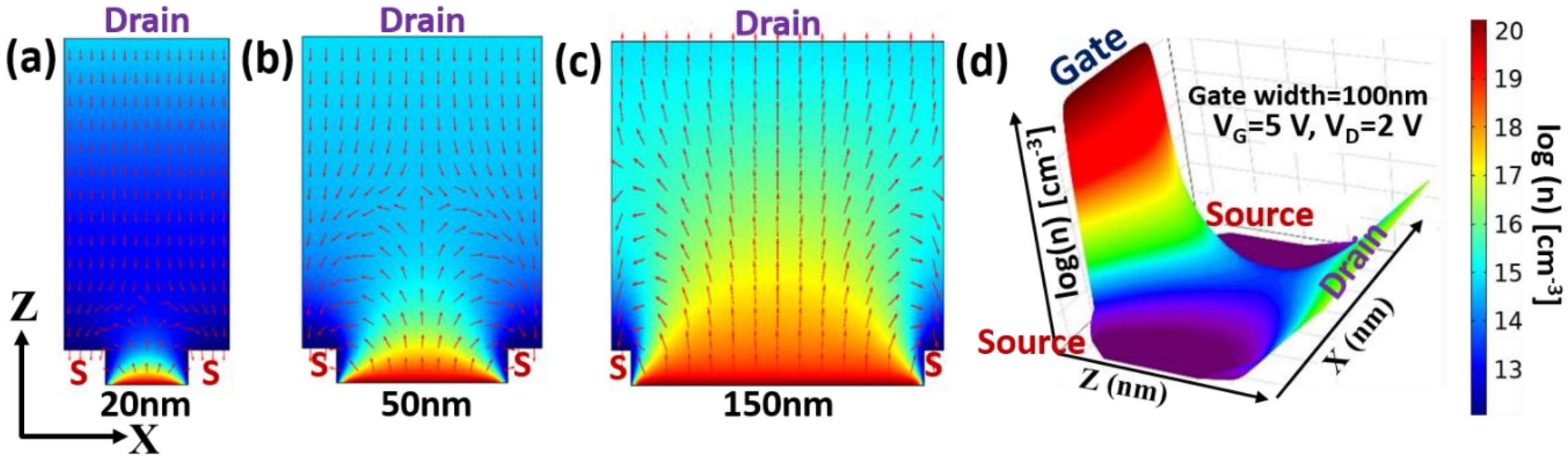

**Figure 3.** Simulated charge carrier concentration (n, represented with colour bar) and field (quiver plot) in VOFET devices with gate openings/width ($w_g$) **(a)** 20 nm, **(b)** 50 nm, and **(c)** 150 nm at $V_G$=5 V, $V_D$=0.1 V. X and Z axis indicate the device's horizontal and vertical length, respectively. **(d)** Charge carrier density distribution (logarithmic scale) along the channel at gate width of 100 nm at $V_G$=5 V, $V_D$=2 V.

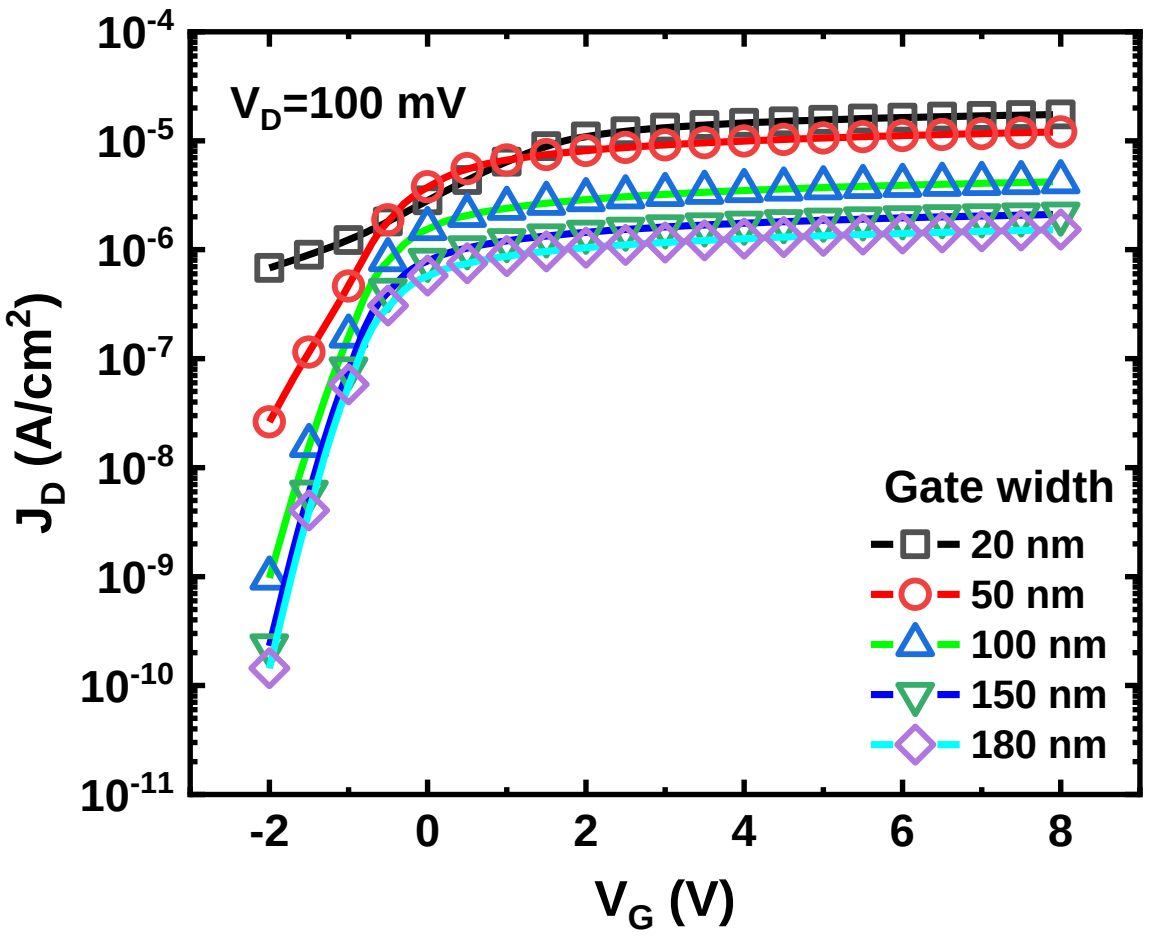

**Figure 4.** Transfer characteristics of VOFETs with various gate openings at $V_D$=0.1V.

The On/Off ratios and SS values are extracted from the transfer characteristics of Figure 4 and are plotted in **Figure 5 (a)** and **(b)** along with FF in the inset of SS in Figure 5 (b). Figure 5 (a) shows that the On/Off ratio initially increases with gate width and later saturates. Here it is also worth noting that the high On/Off ratios in larger gate width is because of the low off-state current in the negative gate voltage direction as seen in Figure 4. This same gate voltage region which is from -2 V to 0 V is also responsible for high SS values which means that for optimized device geometry, VOFET performance can be improved by operating it in the negative gate direction and by increasing the gate perforations sizes w.r.t source size. So, it can be said that the actual working principle of VOFETs is similar to depletion mode MOSFETs where a reverse field is required to switch off the transistor. The results show that the reason behind the high On/Off ratio and low SS in the negative gate voltage direction is the strength of the gate field which decreases the carrier flow from the source to the drain electrode which is shown in **Figure 6**. The high opposite field in the negative gate voltages hinders charge carrier accumulation and transport from source to drain as seen in Figure 6 (a) by repelling charge carrier not only from the gate interface but also from the top of the source electrode which hence decreases the transfer current density (yellow region representing $10^4$ $cm^{-3}$). Figure 6 (b), (c) and (d) show relatively much higher charge carrier concentration, which results in much higher transfer current density. This effect is only visible in devices with higher FF values, the field induced by gate bias reaches easily into the bulk of the OSC and suppresses the drain-source field and hence becomes a deciding factor for the current flowing through it. Simulating the devices with different perforation sizes and fixing the source width (10 nm), a lower limit of the gate width has been achieved, which is 150-200 nm as seen in Figure 5(a), above this gate width value the gate field can penetrate into the OSC and also reduce the current density in negative gate voltages. SS values also show a similar trend with fast initial fall and later saturation at 150 to 200 nm perforation size as seen in Figure 5(b). The FFs of the devices are also calculated as seen in the inset of Figure 5(b) which shows performance enhancement as the gate width increases which fits well with our explanation. All the above observations suggest that a FF dependence which should be always more than 0.8 for achieving better VOFET performance. These results show that the negative gate bias voltage is actually responsible for a higher On/Off ratio and lower SS which demonstrates the operation of VOFETs in reality takes place from negative gate voltage which is similar to depletion mode MOSFET operation.[39]

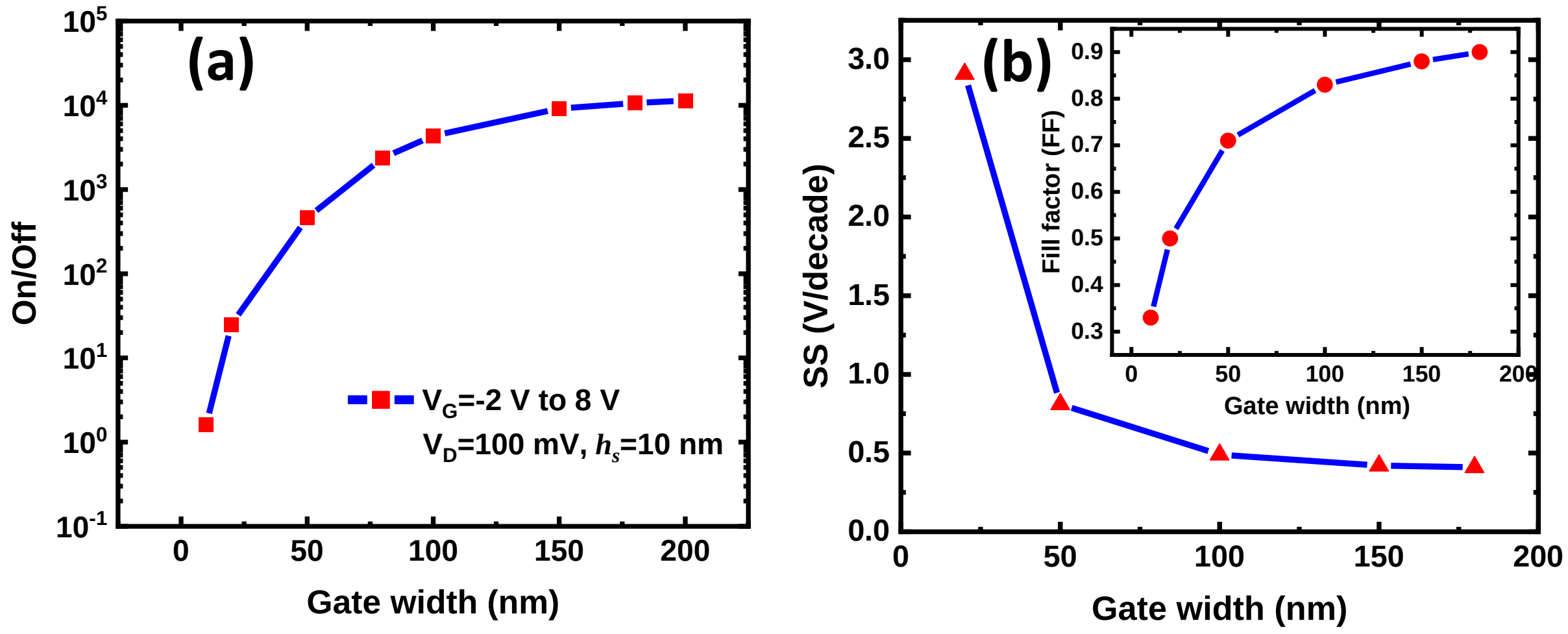

**Figure 5. (a)** On/Off ratio increasing with increasing gate area and saturates after 150 nm. **(b)** The SS value plotted against the device gate width (inset show the FF vs gate width)

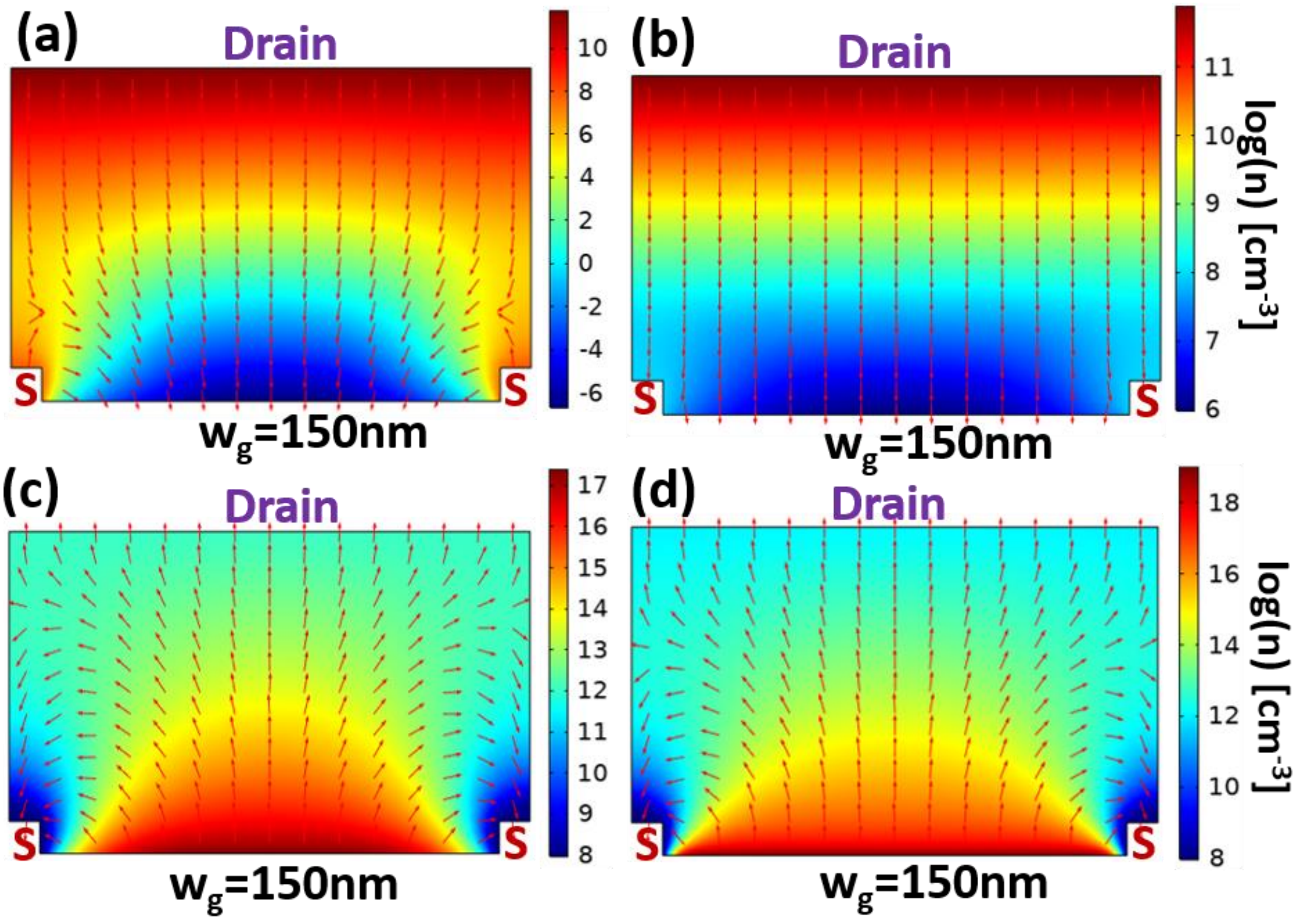

**Figure 6.** Charge carrier concentration (n, represented with colour bar) variation from the source to drain contact along with the changes in electric field inside the channel after applying gate bias, $V_G$ **(a)** -2 V, **(b)** -1 V, **(c)** 0 V, and **(d)** 2 V with gate width, $w_g$ = 150 nm, and drain bias, $V_D$ = 100 mV and source width of 10 nm both sides.

### *2.1.2. Source electrode thickness ($h_s$)*

The source electrode thickness is also a major factor in the performance of the devices because it can control the Off-current. It is seen from the previous section that by choosing the right value of FF, the gate field can further reduce the Off-state current density by repelling charge carriers from the top of the source electrode, which suggests that the effect of the source electrode height can also affect the drain current density. Simulations are carried out with different source electrode thicknesses and by keeping the perforation size the same, somewhere at the better device performance value which is 180 nm. The simulated transfer characteristics plots ($J_D$ vs. $V_G$) for different VOFETs with different source thicknesses of $h_s$ = 10, 20, 50, 70 nm with a fixed gate width $w_g$ = 180 nm, source electrode width $w_s$ **=** 10nm and at $V_D$=0.1 V are shown in **Figure 7(a)**. The Off-state current density is seen rapidly increasing with source electrode thickness which is expected due to the lower thickness of the OSC on top of the source electrode which reduces the charge carrier conduction path from source to drain or vice versa. The on-state current density is also on an increasing trend but with less increment in the values. It can be verified by comparing $h_s$ = 10 nm (black dotted line) and $h_s$ = 70 nm (green dotted line) in **Figure 7(b)** which shows a large difference in $J_D$ values of the Off-state ($V_G$ = -2 V), whereas less increment is seen during the on-state operation of VOFET. The On/Off ratio dependence on source electrode thickness is plotted in **Figure 7(c)** which shows a decreasing trend in On/Off ratios. **Figure 7(d)** shows the SS values with source thickness which is increasing with source electrode thickness suggesting poor device performance at higher source thicknesses. The poor performance at higher source thickness can be explained by **Figure 8** which shows the COMSOL simulation results of the VOFET with a gate width of 50 nm and channel length of 100 nm with three different source heights. The field quiver plot, presented in Figure 8. (a), (b), and (c) are for three different devices having source electrode thickness $h_s$=10 nm, 20 nm, and 50 nm respectively, where for $h_s$=10 nm and 20 nm the gate field accumulation region is well in the channel region, but in case of $h_s$=50 nm gate field is confined by the source electrode thickness. Hence on increasing the source height, the confinement of the gate electric field takes place and its effect on the current is minimized since the accumulated charge carrier cannot participate in the current flow. As the separation between the source and drain becomes less due to higher $h_s$ values, the drain-source bias causes more charge carriers (electrons) to flow from the top surface of the source electrode giving rise to much higher Off-state current values. Here the effectiveness of the negative gate bias is also decreased and hence we see a low On/Off ratio and higher SS in the case of a thick source electrode.

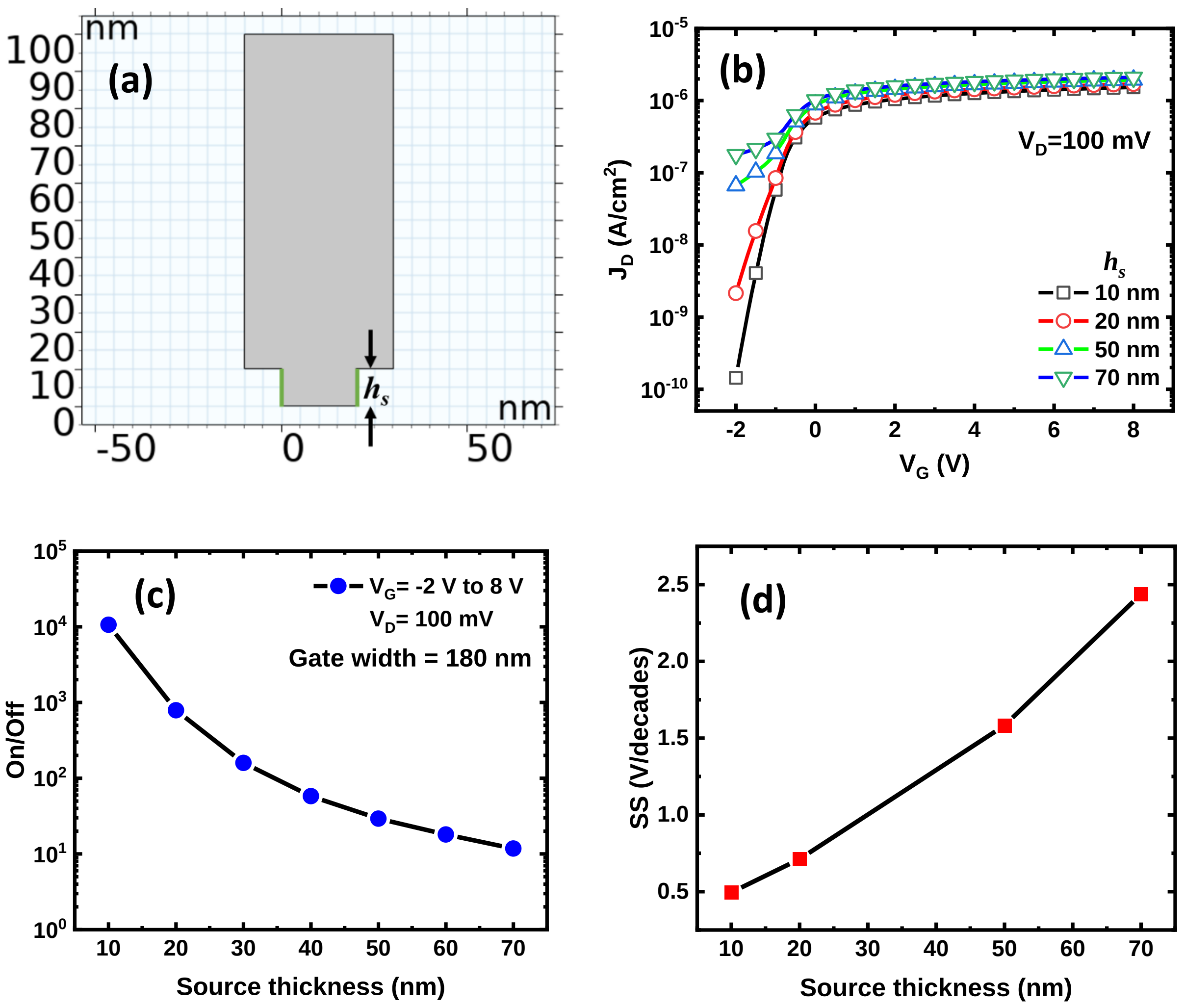

**Figure 7.** **(a)** VOFET simulation model with source thickness $h_s$ = 10 nm. **(b)** Transfer characteristics of devices with various source thicknesses ($h_s$) values at $V_D$=0.1V and gate width=180 nm, **(c)** On/Off ratio vs. source thickness for source perforation/gate width of 180nm. **(d)** *SS* values vs. various source thickness ($h_s$) values.

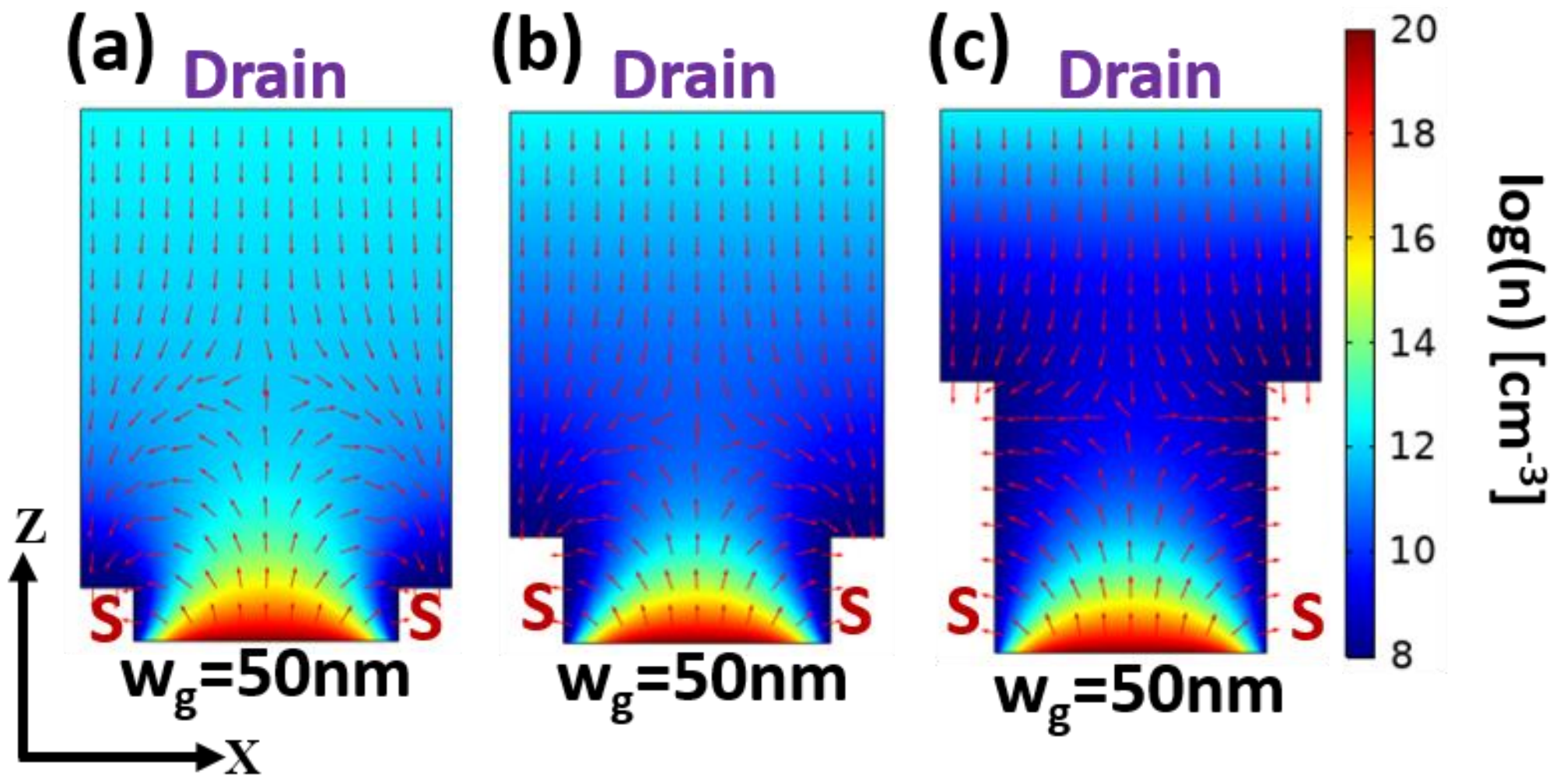

**Figure 8.** Charge carrier density (n, represented with a colour bar) and electric field quiver plot inside the channel of VOFETs at $V_G$=1 V, $V_D$=0.1 V, gate width ($w_g$) of 50 nm and source height, $h_s$ **(a)** 10 nm, **(b)** 20 nm, and **(c)** 50 nm.

*2.1.3. Source electrode width ($w_s$)*

The source width in the device is also responsible for the changes in the current density $J_D$ due to changes in the *FF* as previously seen in the gate width dependence. The model simulated in COMSOL is shown in **Figure 9(a)** for different source widths, $w_s$ with all other parameters kept fixed. The gate width of 180 nm and OSC layer thickness of 100 nm is chosen for all the simulations. The *FF* decreases on increasing the source width ( as the gate width is fixed) which is *FF* = 0.9 at $w_s$ = 10 nm and it is 0.375 when $w_s$=150 nm. The various transfer characteristics are shown in **Figure 9(b)** which shows low Off-state current for the lowest source electrode width. **Figure 9 (c)** shows the On/Off ratio with source electrode width, which is highest ($10^4$) in the case of lowest source electrode width (10nm) and minimum in case of highest source electrode width (150nm). The *SS* value shows a rise with increasing width of the source electrode as seen in **Figure 9(d)**. Comparing the results with the results of the gate width section it can be seen that an optimal value of *FF* is required for the proper operation of VOFETs. The simulated *FF* values at different source widths are shown in **Figure 9(e)**. The dimension of the source width should be such that the gate field can penetrate into the channel and cancel out the drain-source field and hence block the flow of charge carriers when a negative bias is applied at the gate. This condition is required for the low Off-state current and low *SS* values and is also seen in the previous gate width, $w_g$ simulations.

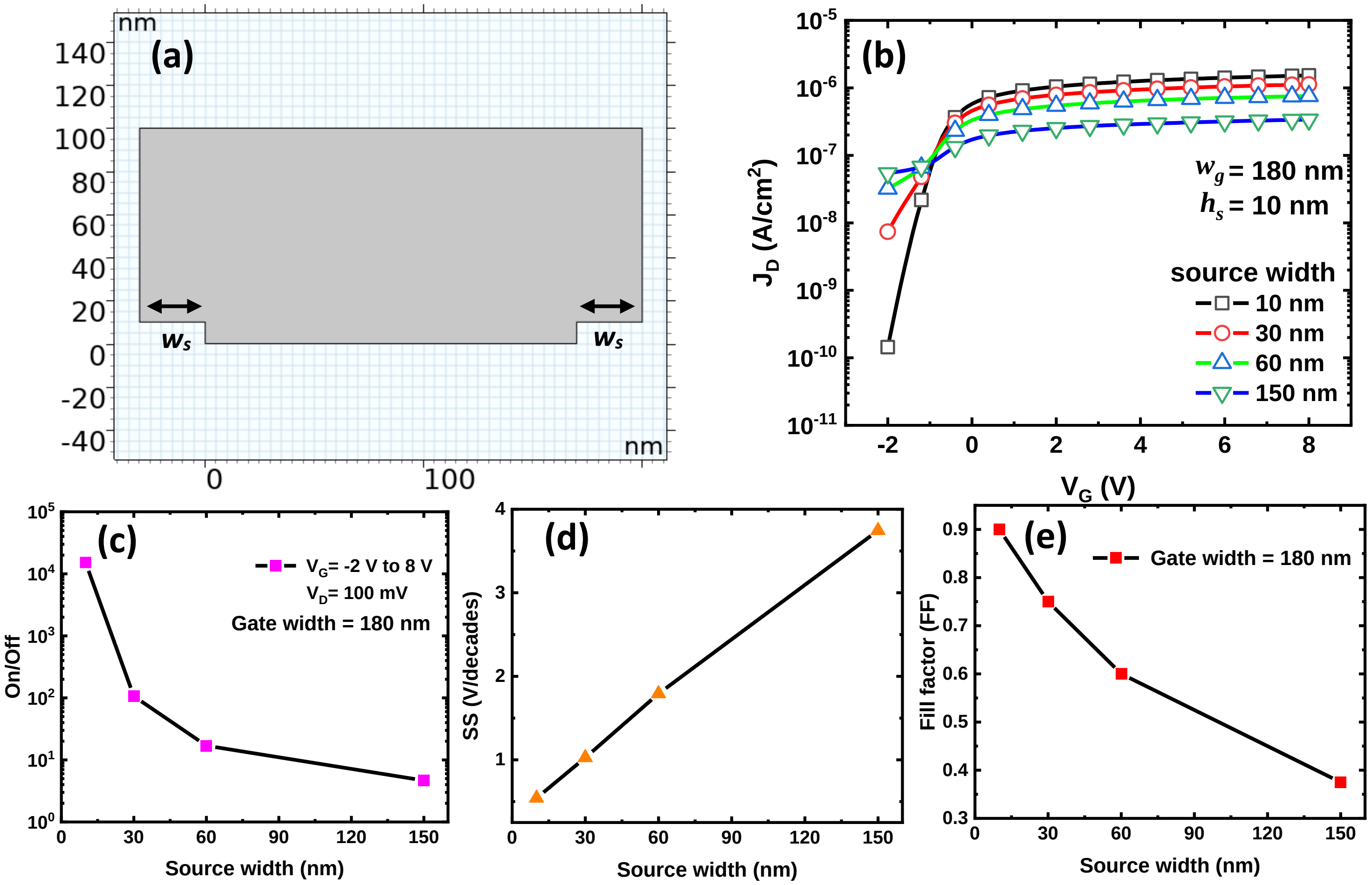

**Figure 9. (a**) VOFET model with source width, $w_s$=30 nm and gate width, $w_g$=180 nm, **(b)** The transfer characteristics $J_D$ vs. $V_G$ response with different values of source widths and at fixed gate width of 180 nm and source height of 10 nm, **(c)** On/Off ratio vs. source width for source perforation/gate opening 180 nm and source thickness 10 nm, **(d)** SS value vs. various source width values, and **(e)** Fill factor vs. source width at gate width=180 nm.

## 2.2. Effect of contact barrier heights

The charge carrier injection barrier at the metal-OSC interface is also an important factor in the operation of VOFETs. Three different source metal electrodes Au, Cu, and Al are simulated with the work functions of 5.0 eV, 4.7 eV, and 4.3 eV[40] respectively. Simulation parameters with good VOFET performance are used from the previous section where a high On/Off ratio and low SS is obtained which is source perforation/gate opening of 180 nm, source height ($h_s$), and source width, $w_s$ of 10 nm. **Figure 10(a)** shows the energy level diagram of the metal contact and OSC, here the metal contact is responsible for charge carrier injection in the semiconductor so the energy barrier plays an important role in injection. Simulations of three different metal contacts are shown in **Figure 10(b)** which clearly shows different device current density $J_D$ for different source metals. The On/Off ratio and SS are also plotted in **Figure 10(c)** and **(d)**, respectively. The charge injection barriers at the source metal-OSC interface for Au, Cu, and Al are approximated to 0.7 eV, 0.4 eV, and 0.22 eV respectively. Larger $J_D$ values can be seen in the case of Al due to the less injection barrier of 0.22 eV, and the highest injection barrier in the case of Au (0.7 eV) results in the lowest $J_D$ as seen in Figure 10 (b). The low injection barrier leads to higher charge carrier injection from the source electrode which can take part in conduction from source to drain, as shown in **Figure 11**, where high charge carrier concentration can be seen at negative gate voltages which results in high current density. Highest On/Off ratio of ~$10^6$ and lowest SS can be seen for the optimum injection barrier of Cu which can be seen in Figures 10 (c) and (d). The simulation results show that both low and high injection barriers can hinder the performance of VOFETs, hence the effect of the injection barrier is also a very important factor in the proper working of VOFETs

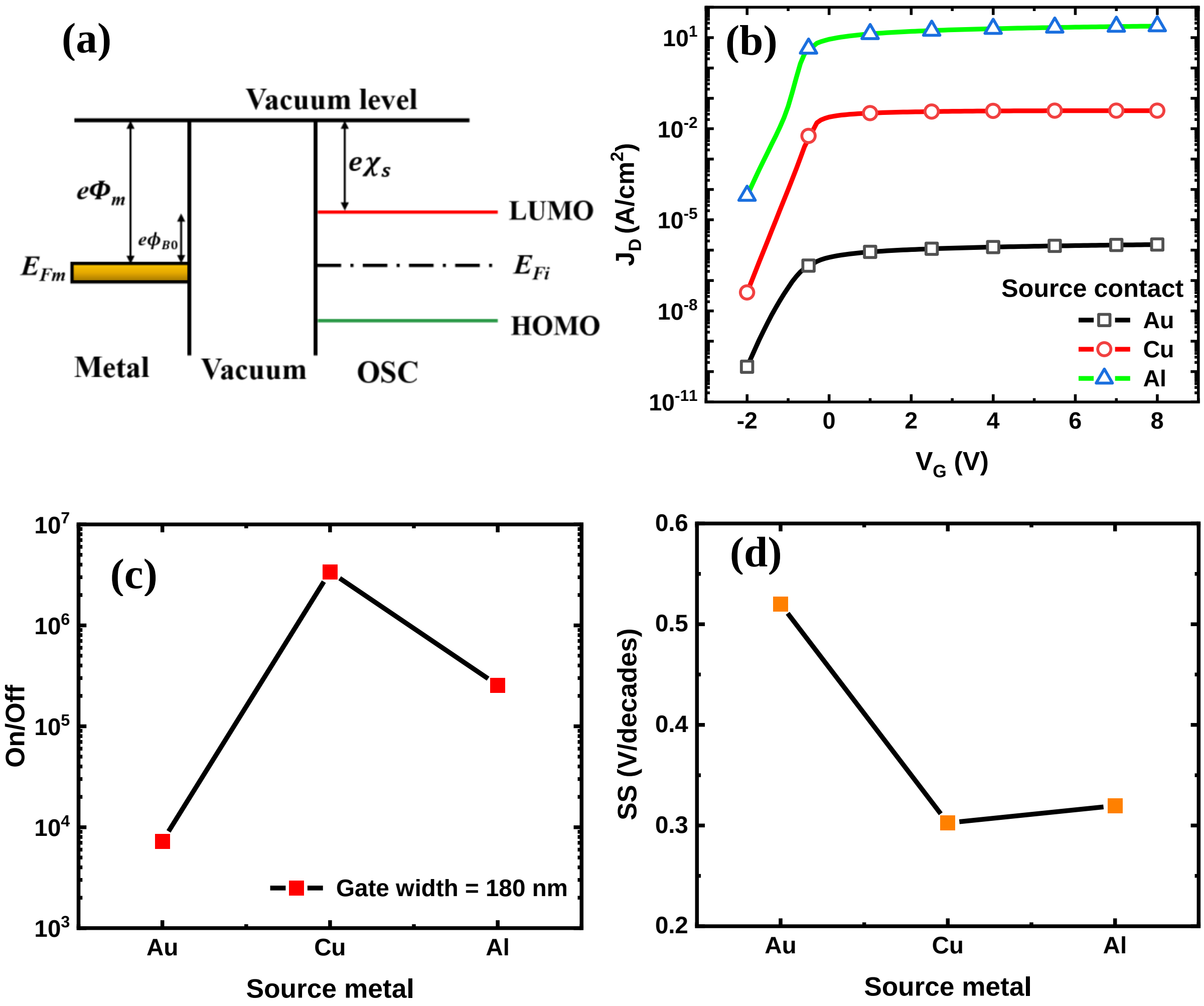

**Figure 10. (a)** Energy band diagram of metal-vacuum-OSC system with metal fermi level between the LUMO and HOMO levels of OSC, **(b)** Transfer characteristics of simulated VOFET with different source electrode, with $w_g$=180 nm, $h_s$=10 nm and $w_s$=10 nm. **(c)** On/Off ratios with source metal, and **(d)** SS changes with source metal.

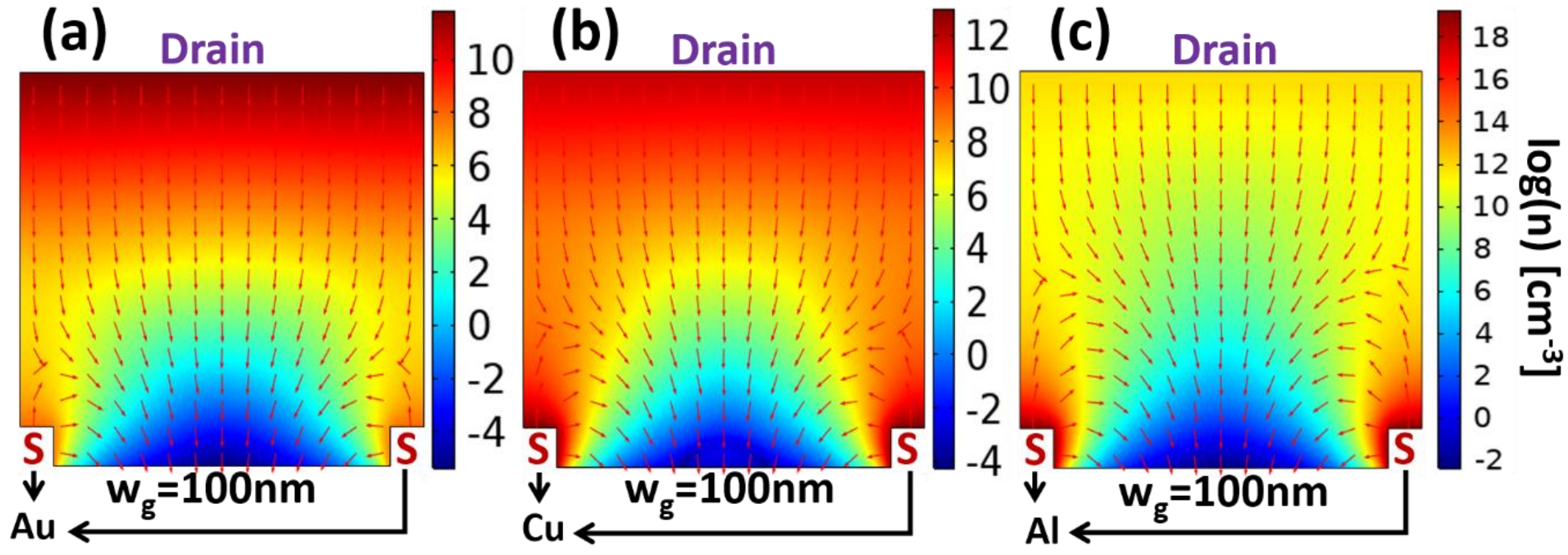

**Figure 11.** Simulated charger carrier density (n, represented as colour bars) and field lines in VOFETs for (a) Au, (b) Cu, and (c) Al source metals at $V_G$ = -2 V, $V_D$ = 0.1 V with the gate width, $w_g$ = 100 nm, and source width, $w_s$ = 10 nm.

### 2.3. Effect of organic semiconductor (OSC) properties

OSC properties can also influence the performance of the VOFETs so in this section effects of free charge carrier concentration/doping density and mobility are simulated.

*2.3.1. Doping density*

The density of dopants or free charge carrier concentration in OSC can also affect the behavior and performance of VOFETs. Simulations are carried out for different charge carrier densities and the results are shown in **Figure 12**, where the charge carrier concentration is varied from $10^{10}$ cm$^{-3}$ to $10^{15}$ cm$^{-3}$. The charge carrier distribution is assumed to be uniform in the whole semiconductor region. Here good device performance-based VOFET architecture is used for simulations with gate width ($w_g$) 180 nm and other parameters like source height, $h_s$, source width, $w_s$ , and thickness of OSC layer are taken as 10 nm, 10 nm, and 100 nm respectively. The source metal is chosen as Au as considered in most of the simulations. Figure 12(a) shows high Off-state current density in the case of high doping concentration while negligible changes in On-state current density. The On/Off ratio and *SS* are shown in Figure 12(b) and (c), respectively which show a decrease in On/Off ratios and an increase in SS at higher concentrations. It can be concluded from the simulations that at higher charge carrier concentration the VOFET device performance degrades, and a lower value of charge carrier concentration is more optimal for device operations due to the rise in the Off-state charge carrier flow from the source to the drain electrode.

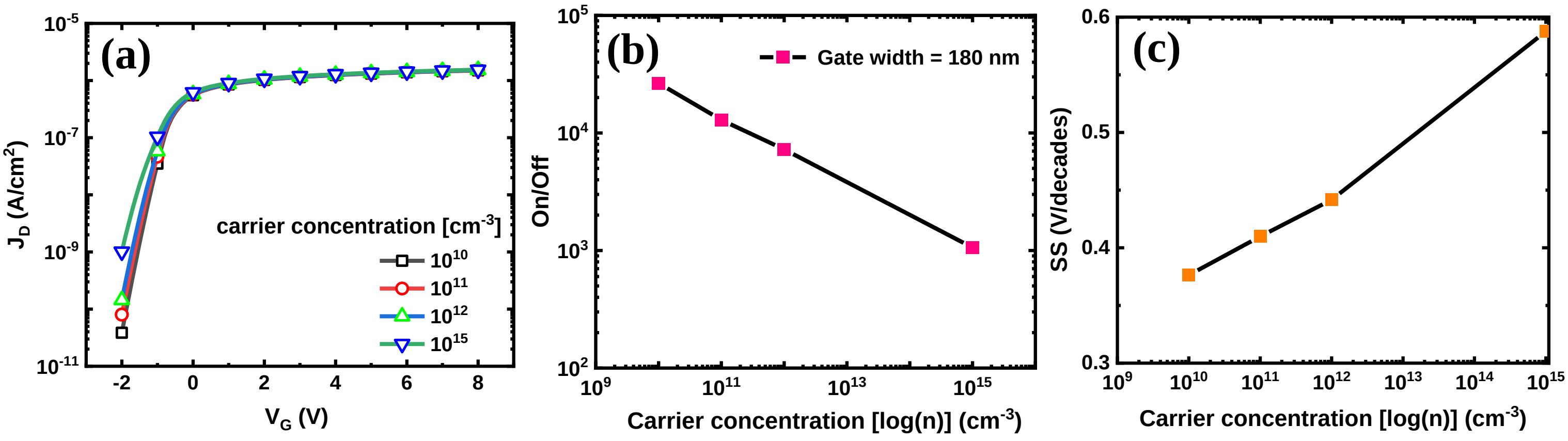

**Figure 12. (a)** Simulated transfer characteristics of VOFET with gate width ($w_g$) 180 nm and other parameters like source height, $h_s$, source width, $w_s$ and thickness of OSC taken as 10 nm, 10 nm and 100 nm, respectively at four different donor density of $10^{10}$ cm$^{-3}$, $10^{11}$ cm$^{-3}$, $10^{12}$ cm$^{-3}$ and $10^{15}$ cm$^{-3}$, **(b)** On/Off ratio vs. charge carrier concentration, and **(c)** SS vs. charge carrier concentration.

*2.3.2. Charge carrier mobility*

Charge carrier mobility is a deciding factor in the value of current density in any OSC-based device, hence different charge carrier mobility simulations are also performed. OSC carrier mobility is varied from 0.1 to $10^{-8}$ cm$^2$ V$^{-1}$ s$^{-1}$ and analyzed in simulation by keeping the structural parameters constant with high On/Off ratio and low *SS* as given in previous sections. The mobility values of electron and hole are assumed constant and independent of any applied field throughout the simulation. Source and drain electrodes are also kept the same in all simulations which are Au and Al respectively. The simulated transfer characteristics are shown in **Figure 13(a)**, which shows higher current density $J_D$ (on and off-states) with higher charge carrier mobility values and the rise is proportional to the rise in the mobility values. The On/Off ratio and SS show almost no changes with an increase in the charge carrier mobility as seen in **Figure 13(b)** and **(c)**, respectively. In conclusion, the effect of mobility is just that it will increase the current density proportional to the increment in mobility value.

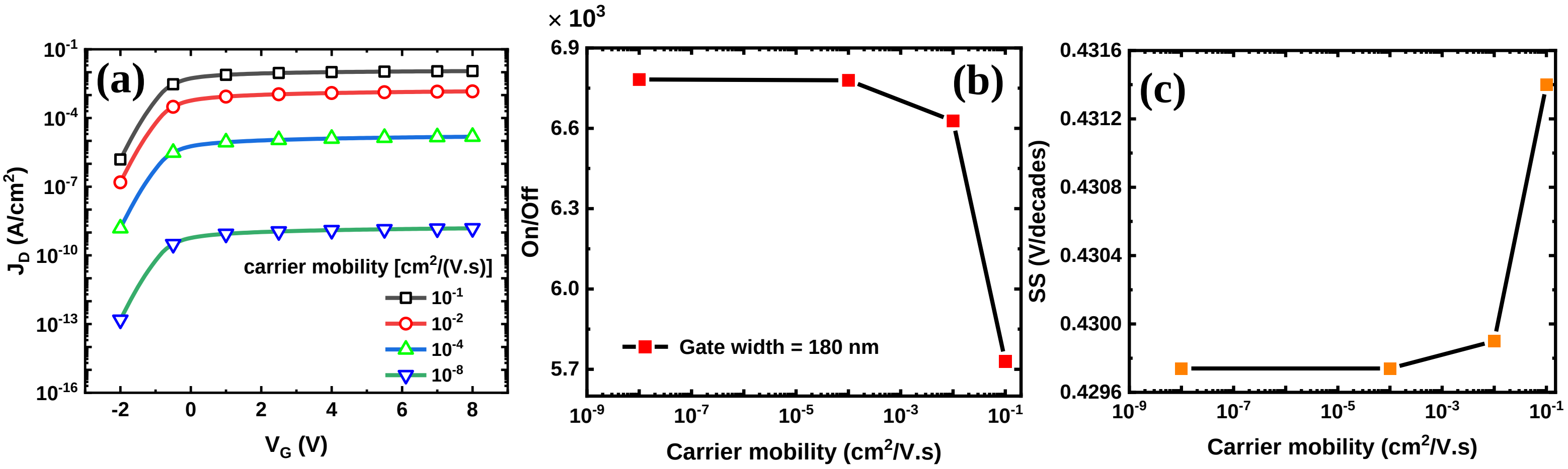

**Figure 13. (a)** Transfer characteristics at different charge carrier mobility values of VOFET with gate width ($w_g$) 180 nm and other parameters like source height, $h_s$, source width, $w_s$ and thickness of OSC taken as 10 nm, 10 nm and 100 nm, respectively. The drain bias is fixed at 0.1 V, **(b)** On/Off ratio vs. carrier mobility, and **(c)** *SS* vs. carrier mobility.

## 3. Conclusion

In this study, we have demonstrated and explained how device architecture, metal injection contacts, and semiconductor properties can be optimized to achieve high-performance VOFETs. The first factor which is a major contributor to the device performance is the device architecture where gate opening, source electrode width, and eight are optimized to get the best VOFETs with low Off-state current. VOFET with gate opening between 150nm, source width 10nm, and source height 10 nm have shown better performance due to the penetration of the gate field in the channel region and also around the source electrode. The penetration of the gate field in the OSC layer and top of the source electrode results in low charge carrier concentrations around the source electrode at negative gate voltages which further reduce the current density and hence improve the On/Off ratio and SS of VOFET. The result can also be understood in terms of the FF value which is the ratio of gate width/area to total width/area with the source electrode, and source thickness value, all three parameters must be optimized to achieve good VOFET performance. The main factor in improving the VOFET performance is the negative gate field which repels the charge carrier from the gate dielectric/semiconductor interface and also around the top of the source electrode. The results also show that VOFETs are similar to depletion mode MOSFET devices which required negative gate voltages to switch it off. The effect of the charge carrier injection barrier is also seen to affect the current density of the device with some enhancement in the device performance by choosing the optimum charge carrier injection barrier. In the end, simulations are done for different values of charge carrier density and mobility in OSC and found that lower values of free charge carriers are more suited for achieving better device performances whereas increasing the charge carrier mobility simply increases the current density of VOFETs without any improvement in the performance.

**Acknowledgments**

The authors would like to thank the Indian Institute of Technology Bombay (IITB) for the seed grant and fellowship.

ToC text:

The Off-state current in Vertical field effect transistors (VOFETs) is minimized by various methods which increases the complexity and processing cost. The simulation studies on device architecture, contact barrier, and semiconductor layer show low off-state current, and high performance can be achieved without the addition of various complex procedures, where the majority of optimization takes place due to device architecture.

Ramesh Singh Bisht, and Pramod Kumar*

**Simulation study of various factors affecting the performance of Vertical Organic Field-Effect Transistors**

ToC figure

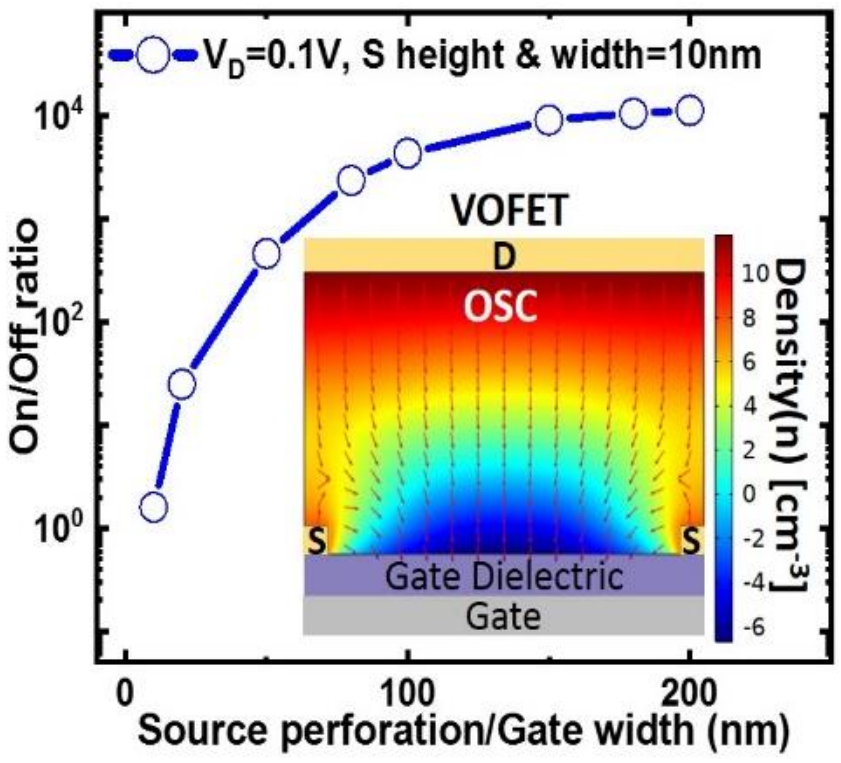